\documentclass[useAMS,usenatbib]{mn2e}

\usepackage{amsmath,texlogos} 
\usepackage{natbib}
\usepackage{subfigure}
\usepackage{graphicx}
\usepackage{txfonts}

\title[A massive exoplanet candidate around KOI-13]{A massive
  exoplanet candidate around KOI-13: Independent confirmation by
  ellipsoidal variations}
\author[D. Mislis]{D. Mislis$^{1}$\thanks{E-mail: misldim@ast.cam.ac.uk}, S. Hodgkin$^{1}$ \\
  $^{1}$Institute of Astronomy, Madingley Road, Cambridge CB3 0HA, UK
}

\begin{document}

\date{Accepted : 2012 February 08}

\pagerange{\pageref{firstpage}--\pageref{lastpage}} \pubyear{2002}

\maketitle

\label{firstpage}

\begin{abstract}
  We present an analysis of the KOI-13.01 candidate exoplanet system
  included in the September 2011 Kepler data release. The
  host star is a known and relatively bright $(m_{\rm KP} = 9.95)$
  visual binary with a separation significantly smaller (0.8 arcsec)
  than the size of a Kepler pixel (4 arcsec per pixel). The Kepler
  light curve shows both primary and secondary eclipses, as well
  as significant out-of-eclipse light curve variations. We confirm that the
  transit occurs round the brighter of the two stars. We model the
  relative contributions from (i) thermal emission from the companion,
  (ii) planetary reflected light, (iii) doppler beaming, and (iv)
  ellipsoidal variations in the host-star arising from the tidal
  distortion of the host star by its companion. Our analysis, based on
  the light curve alone, enables us to constrain the mass of the KOI-13.01
  companion to be $M_{\rm C} = 8.3 \pm 1.25M_{\rm J}$ and thus
  demonstrates that the transiting companion is a planet (rather than
  a brown dwarf which was recently proposed by \cite{b7}). The high
  temperature of the host star (Spectral Type A5-7V, $T_{\rm eff} =
  8511-8020$ K), combined with the proximity of its companion
  KOI-13.01, may make it one of the hottest exoplanets known, with a
  detectable thermal contribution to the light curve even in the Kepler optical
  passband.  However, the single passband of the Kepler light curve does not
  enable us to unambiguously distinguish between the thermal and
  reflected components of the planetary emission. Infrared
  observations may help to break the degeneracy, while radial velocity
  follow-up with $\sigma \sim$ 100 m s$^{-1}$ precision should confirm
  the mass of the planet.
\end{abstract}


\section{Introduction}

In Mislis et al. (2011) we demonstrated the potential for estimating
planetary masses solely from light curve (LC) data for a restricted
sample of systems. The distinguishing characteristic of these systems
is that the planet must be both massive enough and close enough to the
host star to induce significant tidal distortions in the stellar
ellipsoid. Such non-sphericity in the plane of the sky for a rotating
star will lead to a periodic photometric signal which is detectable
given sufficient signal-to-noise. In Mislis et al. (2011) we applied
the technique to the transiting exoplanet HAT-P-7 observed with
Kepler, and estimated the mass of the planet to be
$M_p(LC)=1.27\pm0.12M_J$ which is very close to the RV solution,
$M_p(RV)=1.20\pm0.05$ (Welsh et al. 2010).

The second exoplanet candidate list released by the Kepler team in
February 2011, and published in \citep{b1}, contains 1235 exoplanet
candidates. The light curves span measurements taken between May 2009 and
September 2011. Motivated by our previous work, we searched the light curves of all
1235 planetary candidates for evidence of ellipsoidal variation. The
brightest and best candidate for a star showing strong out-of-eclipse
ellipsoidal variations is KOI-13.01\footnote{Shortly before submission
  of this manuscript we became aware of an analysis of the same
  system, drawing similar conclusions, submitted by Mazeh et
  al. (2011arXiv1110.3512).}  candidate, which is identified with an
A5-7V type star (with a close companion 0.8 arcseconds to the
west\footnote{Szabo et al. 2011}). The light curve magnitude is $m_{\rm KP}$ =
9.95 (Kepler passband), and shows primary and secondary eclipses with
a published period of 1.7635892 days. The shape of the eclipses are
indicative of either a small transiting companion, or a blended
system, whereby a deeper eclipse is diluted by contaminating light
from nearby stars in the same Kepler pixels.

KOI-13.01 is already the subject of a through analysis of the source
characteristics and the transit shape by Szabo et al. (2011). They
conclude that the system comprises a double A-star wide binary system,
the more massive component of which is being eclipsed by a brown-dwarf
or very-low-mass star on the basis of a radius determination $R_{\rm
  C}=2.2\pm0.3R_{\rm J}$ ($R_{\rm C}$: companion radius). In this
paper we independently confirm that the brighter component of the
visual binary is the host star for the eclipses
(Section~\ref{data}). Barnes et al. (2011) use a more sophisticated
analysis to measure the radius of the host star and transiting
companion, including the effects of a gravity-darkened
rapidly-rotating host star. The updated values ($R_{\rm C} = 1.4 \pm
0.016 R_{\rm J}$, $R_{\star}= 1.77 \pm 0.014 R_{\odot}$) are
significantly smaller than those found by Szabo et al. 2011, and lead
Barnes et al. (2011) to conclude that KOI-13.01 may be a planet. 
Finally, Rowe et al. (2011) presented the system at the 2011 AAS
  meeting and also suggested the companion is a planet.

In Section~\ref{modelling} we perform modeling of the
out-of-eclipse photometric variation, which puts very strong
constraints on the mass of the eclipsing/transiting companion, and
lends support to the argument that KOI13.01 has a planetary mass.

\section{Kepler Data}
\label{data}

\begin{figure}
 \includegraphics[width=80mm]{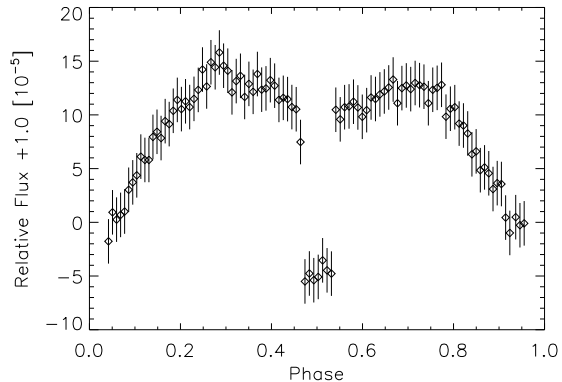} 
 \caption{Phase folded light curve and the secondary eclipse of KOI-13
   system. The primary transit has been removed to enhance the
   visibility of the out-of-transit variability.}
 \label{all}
\end{figure}

We have used the Kepler public light curve and centroid curve data available
from the Kepler archive hosted by the Multimission Archive at 
STScI (MAST)\footnote{http://archive.stsci.edu} for our
analysis. The observations comprised four long-cadence and four
short-cadence datasets (spanning Kepler Q1--Q3). The data were
  processed by the Pre-Search Data Conditioning (PDC) module (Kepler
  Data Processing Handbook)\footnote{available via
    http://keplergo.arc.nasa.gov/Documentation.shtml}. For objects
  brighter than $m_{\rm KP} = 11.3$ some saturation, and then bleeding,
  of the CCD pixels will occur. This charge is not lost, and using an
  appropriately shaped mask will preserve the flux, and maintain
  precise photometry even for extremely bright sources (Gilliland et
  al. 2010, 2011). Both short-cadence and long-cadence data are
  equally affected given that they use the same 6 second exposure.
Figure~\ref{all} we show the Kepler light curve folded on the period
of 1.7635892 days found by Borucki et al. (2011), with phase zero set
to be the center of the primary minimum.

\subsection{Identification of the companion host star}

Borucki et al. (2011) suggest that the source associated with KOI-13
is actually a double star (separation ~0.8 arcsec, $\Delta
mag=0.4$). Szabo et al. (2011) show that the Kepler source is
coincident with the double star (BD+46~2629A and B, hereafter KOI-13A
\& KOI-13B), two stars of spectral class A, separated by 1.18$\pm0.02$
arcseconds with a magnitude difference of $\Delta mag=0.20 \pm 0.04$
magnitudes.

Howell et al. (2011) use speckle imaging to find a separation of 1.16
arcseconds and $\Delta mag=0.85$ at 562nm ($\Delta mag=1.03$ at 692nm)
between the two components. However, this photometric difference
between the two stars is in serious disagreement with the Szabo et
al. (2011) photometry.  Having inspected the Szabo et al. (2011)
image, we have opted to adopt their value for the magnitude difference
between the two stars in the Kepler bandpass (as do Barnes et
al. 2011). We note that this is a V-band measurement, which is not
identical to the Kepler passband, but for an A-type star, the
differences are small.  In our analysis (Section 5) we consider the
effects of changing the contribution of this companion star to the
Kepler light curve. Based on an unresolved spectrum, and a combination of
resolved and unresolved photometry, Szabo et al. use model isochrones
to find effective temperatures for the A and B components of $T^{\rm
  A}_{\rm eff}=$ 8511K (8128--8913K) and $T^{\rm B}_{\rm eff}=$ 8222K
(7852--8610K) respectively.

\begin{figure}
 \includegraphics[width=80mm]{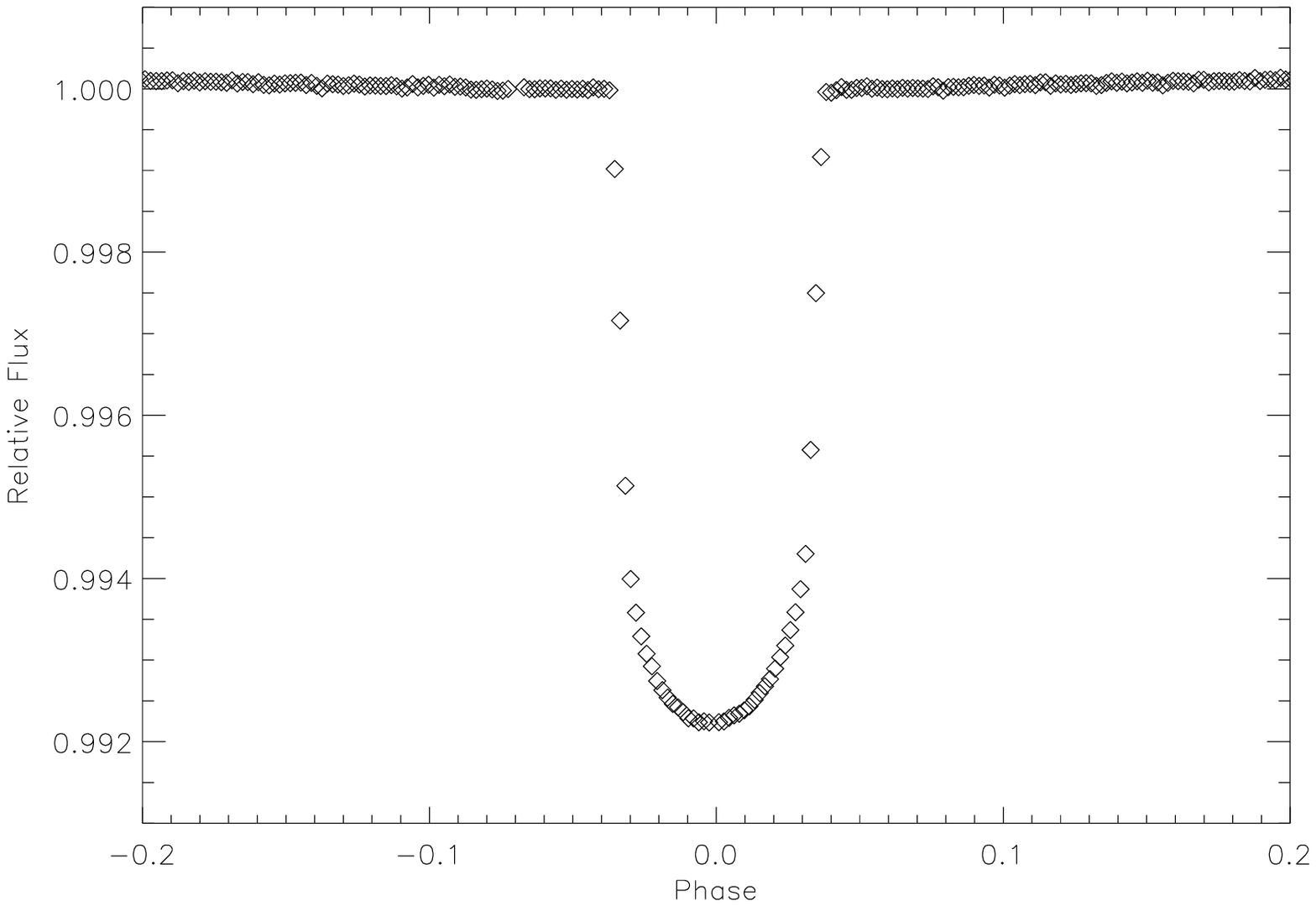} \\
 \includegraphics[width=80mm]{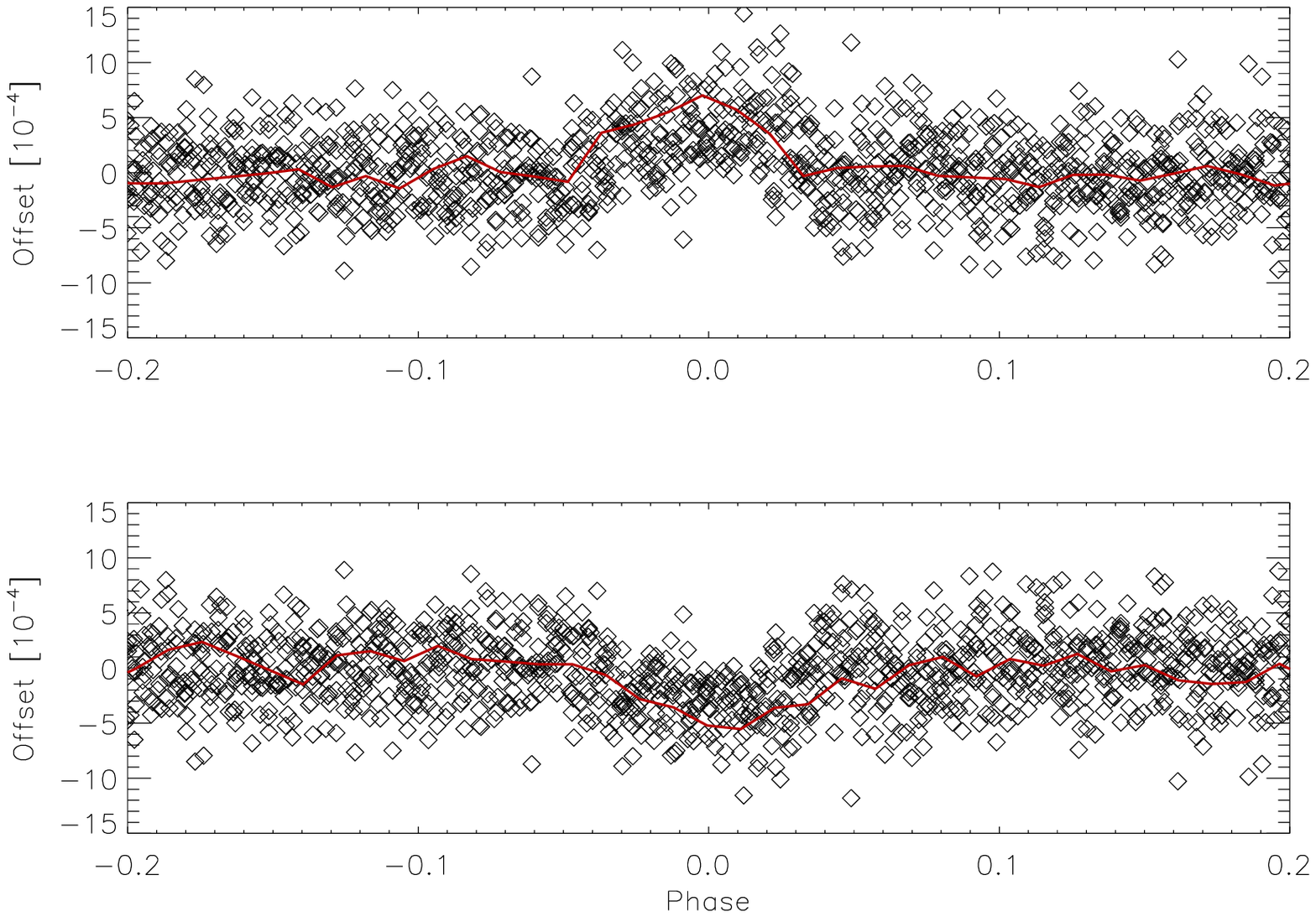} \\
 \includegraphics[width=80mm]{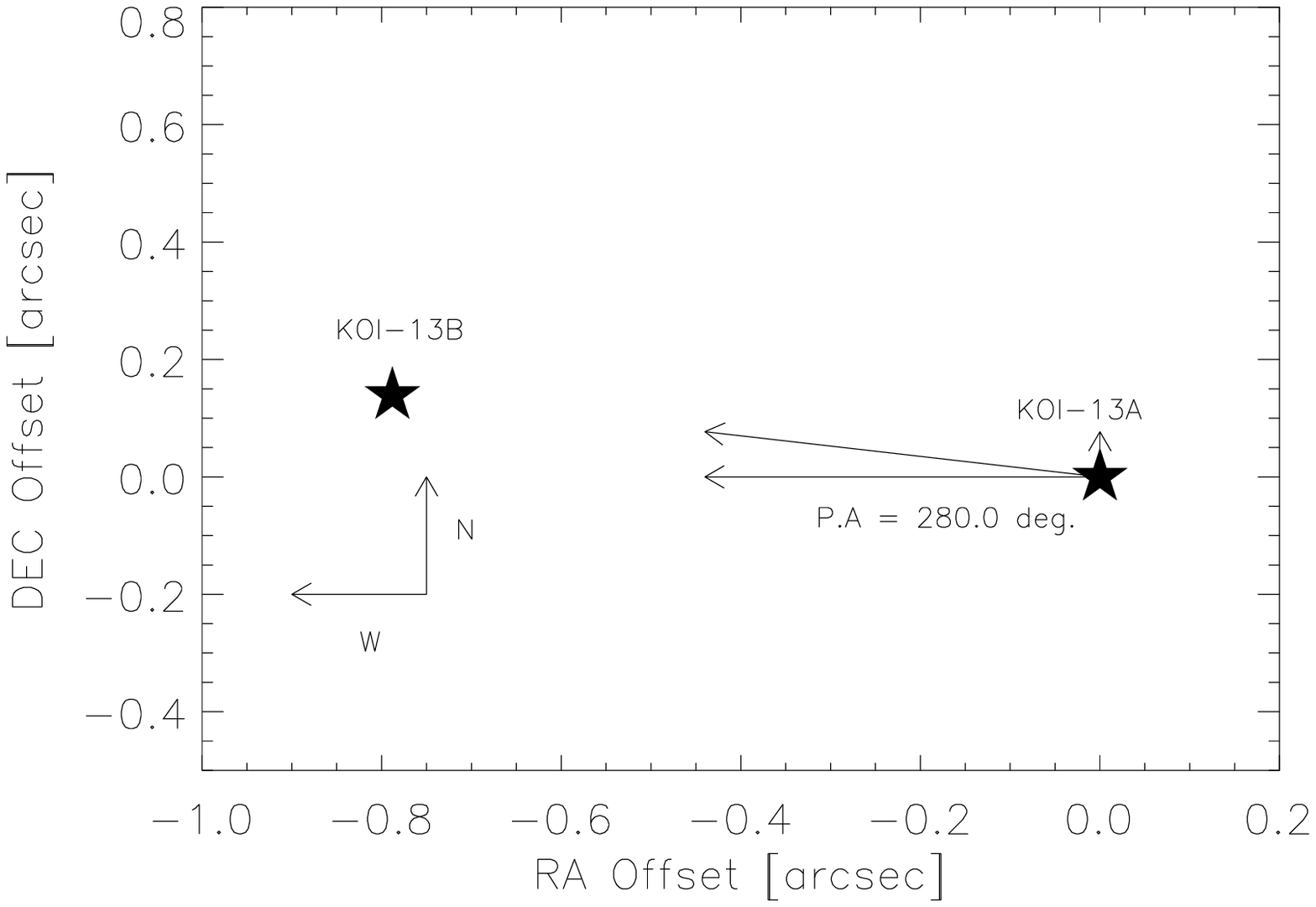} 
 \caption{\textbf{Top:} The transit phase folded light
   curve. \textbf{Middle:} X \& Y axis offset of the centroid. The
   offset is periodic (same period as the transit) in the same
   phase. Offset variations show that the primary star becomes fainter
   during each transit event. \textbf{Bottom:} RA \& Dec coordinates of the KOI-13A 
(right) and KOI-13B (left) plus the RA \& Dec offset vectors. 
The (0,0) point refers to the KOI-13A coordinates ($RA=286.971^{o}$
 \& $Dec=46.8684^{o}$). The vectors have been magnified by a factor of 100. }
 \label{lightcurveandcentroid}
\end{figure}

The centroid curves for KOI-13 are shown in the middle panel of
Figure~\ref{lightcurveandcentroid}. Batalha et al. (2010) have
demonstrated that the very tiny shifts in centroid, expected for
blended occulting sources, can be measured accurately from the Kepler
data. Thus they can be used to show where the largest variations in
flux are occurring with respect to the center-of-light. The strong
periodic offset in the Kepler centroid curve data agrees extremely
well with the time of primary
transit. Figure~\ref{lightcurveandcentroid} (bottom) also shows an
illustration of the direction (and magnified amplitude) of the source
centroid during primary eclipse. This can be explained simply, if
KOI-13A becomes fainter during every transit and the center-of-light
moves towards the west (towards KOI-13B). This interpretation
independently confirms the conclusions from Szabo et al. (2011) who
used both the Kepler pixel light curves, and additional ground-based lucky
imaging. Thus we confirm that KOI-13A is the host to the companion.

\subsection{Removal of the blended light}
\label{radius}

For our analysis, the Q1, Q2 and Q3 datasets were used, corresponding
to a light curve spanning 418 days in total. The light curve was first corrected to
remove the flux from the secondary which we assume to be constant. A
magnitude difference of $\Delta mag = 0.2$ was used, and 45 \% of the
total flux is subtracted from the Kepler light curve \citep{b7}.

The light curve was phase-folded on the best-fit Kepler period of P=1.7635892
\citep{b1}, rebinned by a factor 1770, and phase=0 set to be the
midpoint of the primary transit. We use the updated values for $R_{\rm
  C}$, $R_{\star}$, $\alpha$ (semi-major axis of the orbit) and
inclination ($i$) from Barnes et al. 2011 ($R_{\star}=1.76 \pm 0.014$
$R_{\odot}$, $R_{\rm C}=1.4 \pm 0.014$ $R_{\rm J}$, $\alpha$ = 0.0367
AU \& i=85.9). The phase-folded light curve, including the secondary eclipse
(but excluding the primary transit to emphasize the out-of-eclipse
variation), is shown in Fig. \ref{all}. Finally we found no evidence
for additional periods in the Kepler corrected light curve, such as Mazeh et
al. 2011 suggested (P = 1.0595 days).

\begin{figure}
 \includegraphics[width=80mm]{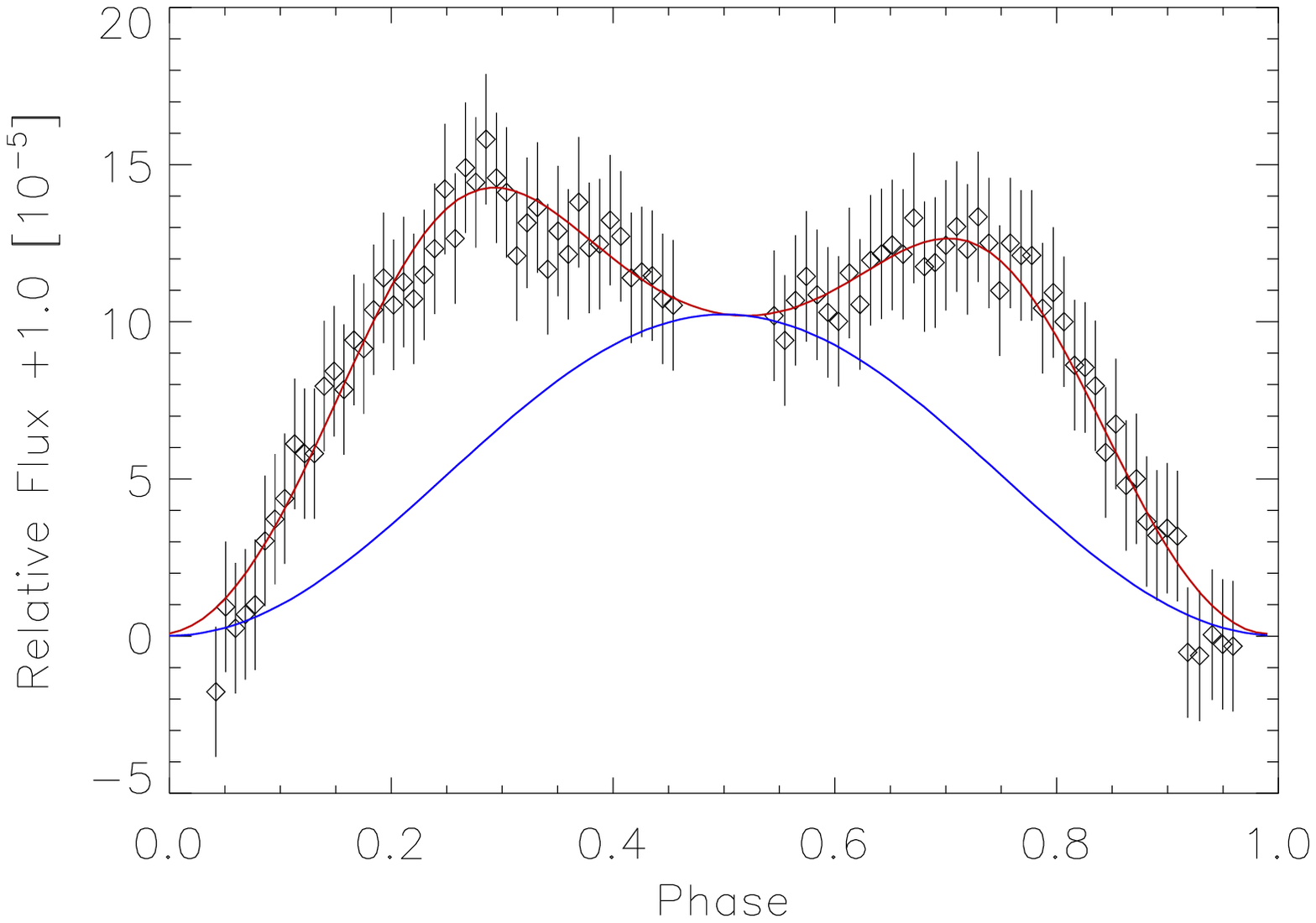} \\
 \includegraphics[width=80mm]{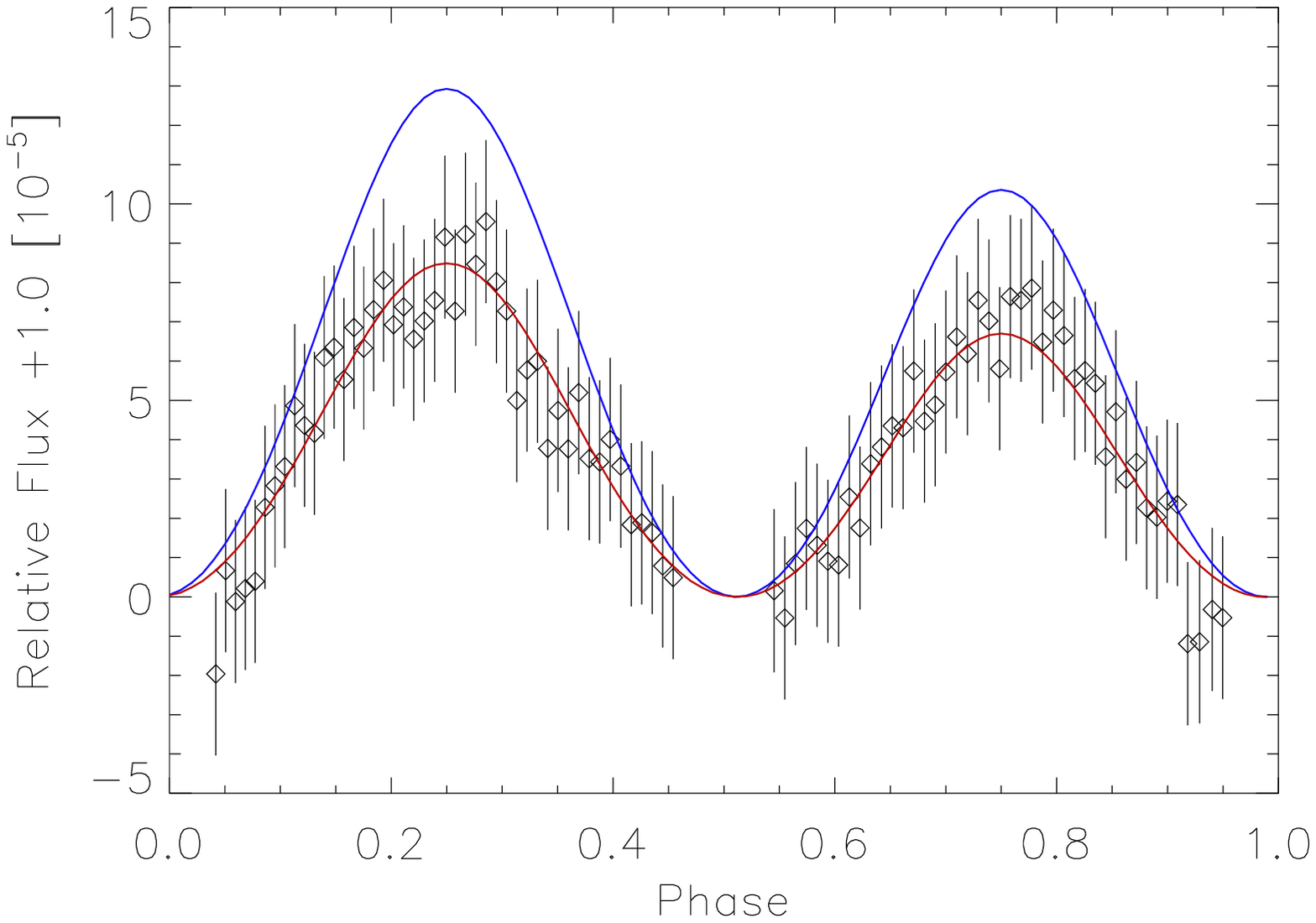} 
 \caption{\textbf{Top:}
 Phase folded light curve and the best fit model ($model-3$) including all
 three components ($f_{\rm th}$, $f_{\rm ref}$, $f_{\rm ell}, f_{dop}$). The
 blue solid line refers to the $f_{\rm th}$+$f_{\rm ref}$ component. The additional flux from ellipsoidal variations at phase $a=0.5$ is zero, 
that's why the total flux at the same phase becomes only from the
$f_{\rm th}$+$f_{\rm ref}$ component (ignoring the secondary eclipse). \textbf{Bottom:} 
The ellipsoidal variation residuals and the best fit (red solid
line). The mass of the body which causes these variations must be
$M_{\rm C} \sim 8.3$ $M_{\rm J}$. The blue solid 
line refers to a brown dwarf (BD) ellipsoidal model with mass $M_{\rm BD} =
13.0M_{\rm J}$. }
 \label{models}
\end{figure}

\section{Modeling}
\label{modelling}

In our modeling, we consider four main components which contribute to
the out-of-eclipse phase dependence of the light curve of KOI-13, summarised in
Equation~\ref{eqtot}.

 \begin{equation}
\centering
\frac{f_{tot}}{f_\star} = 1+\frac{f_{th,day}+f_{th,night}+f_{ref}+f_{ell}+f_{dop}}{f_{\star}}
  \label{eqtot}
\end{equation}
\noindent
In our simple model we include (1) thermal
emission from the companion (day and night side - $f_{th,day}$, $f_{th,night}$ ), (2) reflected light from the surface of the companion, 
(3) ellipsoidal variations due to tidal
forces between the unseeen companion and the star, and (4) flux
variations arising from the Doppler shift (Doppler boosting) of the
stellar spectral energy distribution with respect to the bandpass of
the instrument (Loeb \& Gaudi 2003, Mazeh \& Faigler 2010). This
final component contains no additional parameters. The first four
components are shown in equations 2--5 (for their derivation, and a
detailed discussion, see Mislis et al. (2011)). In Mislis et
al. (2011), we show that if we can disentangle the ellipsoidal
variations from the close companion's thermal and reflected light
components, we can solve for the mass of the companion. \par

 \begin{equation}
\centering
\frac{f_{\rm th}(\theta)}{f_\star} =  A(\theta) \left( \frac{R_{\rm C}}{R_{\star}}
\right)^{2} \cdot Const  
\label{eq01} 
\end{equation}

 \begin{equation}
\centering
\frac{f_{\rm ref}(\theta)}{f_\star} =  \Phi(\theta) \left( \frac{R_{\rm C}}{\alpha}
\right)^{2} \cdot a_{\rm g}
  \label{eq02}
\end{equation}

 \begin{equation}
\centering
\frac{f_{\rm ell}(\theta)}{f_\star} =  \beta \frac{M_{\rm C}}{M_{\star}} \left(
  \frac{R_{\star}}{\alpha} \right)^{3}\sin^{3}(i)
  \label{eq03}
\end{equation}
\noindent

 \begin{equation}
\centering
\frac{f_{\rm dop}(\theta)}{f_\star} =  (3-\rho) \frac{K}{c} 
  \label{eq04}
\end{equation}
\noindent
where $f_{\rm th}/f_\star$ is the relative flux from the thermal
emission (for the day side, $f_{\rm th, day}/f_\star$,
  $A(\theta) = \Phi(\theta)$ and for the night side, $f_{\rm th,
    night}/f_\star$, $A(\theta) = 1-\Phi(\theta)$), $f_{\rm
  ref}/f_\star$ is the relative flux from the reflected light, $f_{\rm
  ell}/f_\star$ is the relative flux from the ellipsoidal variations
and $f_{dop}/f_\star$ is the Doppler Boosting relative
flux. $\Phi(\theta)$ is the phase function, $\alpha$ is the semi-major
axis, $Const(T_{\rm eff}, \lambda, a_{\rm bol}, \epsilon)$ is a
constant function of the effective temperature of the star and the
temperature of the planet (Mislis et al. 2011), and $\beta$ is the
gravity darkening. Finally, $K$ is the radial velocity amplitude, $\rho$ is a function of $\rho(\lambda,T_{eff})$.
 \begin{equation}
\centering
\rho = \frac{e^{hc/k\lambda T_{eff}}(3-hc/k\lambda T_{eff})-3}{e^{hc/k\lambda T_{eff}}-1} 
  \label{eq05}
\end{equation}
\noindent
Our model for the flux changes arising from the ellipsoidal distortion
(Eq. \ref{eq03}) is a simple approximation, and assumes that the
system is in tidal equilibrium (Rowe et al. 2010, Carter et al. 2011,
Mislis et al. 2011). The $a_{\rm bol}$ and $\epsilon$ parameters refer
to the bolometric albedo and the energy circulation respectively. Both
equations \ref{eq01} \& \ref{eq02} are a function of $\Phi(\theta)$, 
but the mass of the secondary body is a dependent parameter
  only in eq. \ref{eq03} \& \ref{eq04}. We note that in our equations
  the terms $R_{\star}/ \alpha$ and $R_{\rm C}/ \alpha$ both
  appear. These are measured directly from the depth and duration of the primary
  transit and the period of the system. For consistancy these values are taken 
from Barnes et al. 2011.

If we were to solve for and remove the first two components (thermal
emission \& reflected light), the residual light curve will contains the mass
information (Fig. \ref{models} - bottom). An important result of the
model is that the ellipsoidal component contributes zero to the total
light curve flux (assuming e=0) at phase=0.5 (centre of the secondary eclipse),
and that the phase function depends only very slightly on the balance
between the day- and night-side properties of any companion's emission
(i.e. it is extremely difficult to measure this with single passband
optical data). Thus the thermal emission and the reflected light
components will be very difficult to distinguish in a single
filter. Nevertheless, one of the goals of modeling the full Kepler light curve 
for KOI-13 is to try to constrain the brightness temperature of
KOI-13.01. The main different between our model and other algorithms
(such as BEER - Faigler \& Mazeh 2011) are that (i) we investigate two
phase functions (Lambert and geometrical spheres - Mislis et al. 2011)
and (ii) we include distinct reflected and thermal components
(including day \& night side, $a_{\rm bol}$, $\epsilon$).  The choice
of phase functions can alter the mass of the planet by 36\%
(see Section 5).

KOI-13A is so hot ($T_{\rm eff} = 8511 \pm 400$ K), and the candidate
planet so close ($\alpha=0.0367$ AU, $\sim$ 4.5$R_{\star}$), that the thermal flux of the
candidate is likely to be non-negligible, even at optical
wavelengths. The equilibrium temperature for any zero-albedo companion
to KOI-13.01 is $T_{\rm eq} = T_{eff}\sqrt{R_{*}/2\alpha} = 2864$K
(for ($R_*=1.694 \pm 0.013$ $R_{\odot}$, Barnes et al. 2011), assuming
that it does not have its own internal source of heating (as one would
expect for a brown dwarf).  The hottest known planet discovered to
date is probably WASP-33b, with an equilibrium temperature of 2750K,
and a brightness temperature of $T_{\rm B} = 3620^{+200}_{-250}$K
(\cite{b12}), measured from ground-based infrared ($\lambda = 0.9\mu
m$) observations of the secondary eclipse (depth of 0.109 per
cent). The relative thermal emission for WASP-33b in the optical
(Kepler passband) would be $~10^{-4} \times f_{\rm *}$ (assuming a
blackbody). We note that the secondary eclipse of KOI-13.01 is $1.2
\cdot 10^{-4} \times f_{\rm *}$ (Szabo et al. 2011) in the Kepler
passband.

As a starting point, we consider three simple scenarios for the phase
light curve data alone, i.e. excluding all datapoints within the primary
transit or the secondary eclipse. In case $C_{therm}$ we assume that
the bolometric albedo of the companion is fixed at $a_{\rm bol}=1.0$,
which means that the planet is perfectly reflective, that the thermal
emission of the planet is zero, and therefore the model contains only
reflected light and ellipsoidal variations. In case $C_{refl}$, we
assume that the geometric albedo is zero $(a_{\rm g}=0.0)$, so the
reflected light is zero and the model contains only thermal emission 
ellipsoidal variations, and Doppler boosting ($f_{dop}$ is a function of $\rho(\lambda, T_{eff})$ and $K$ - Eq. \ref{eq05}). 
Finally, in case $C_{hybr}$ we consider a
hybrid case, and allow $a_{\rm g}$ and $a_{\rm bol}$ to be free
parameters, i.e. both thermal emission and reflected light are present
in the light curve in addition to any ellipsoidal variations. So in $C_{refl}$
we fit for $a_{\rm g}$ and $M_{\rm C}$, in $C_{therm}$ we fit $a_{\rm
  bol}$, $\epsilon$ (energy circulation) and $M_{\rm C}$, and in
$C_{hybr}$ we fit all the parameters above. Eccentricity ($e$) and the
periastron angle ($\omega$) are free parameters in all three
scenarios.

As a final step, we add the primary transit and secondary eclipse data
back into the light-curve, and re-assess $C_{hybr}$ (and name it
$C_{hybr}^b$), primarily to see if the additional information on the
depth of the secondary eclipse can provide useful constraints on the
thermal emission form the planet (as has been suggested by Mazeh et
al. 2011).

An important and unresolved factor is how we treat the phase function 
of the reflected light. We consider two approaches: (1) modelling the planet 
as a Lambert sphere (Russell 1916), assuming that the intensity of the 
reflected light is $f_{ref} = 1/\pi$ at at phase $z = \pi/2$ (as in the case of Venus); 
(2) An alternative choice would be to assume that the reflected light is 
$f_{ref} = 0.5$ at phase $z = \pi/2$. In Mislis et al. (2011) we demonstrate that 
this choice leads to a significantly different phase-contribution for the 
reflected light, and therefore has a significant impact on the derived 
mass of the companion. The companion mass is significantly larger 
(by 36 per cent) if we use the Lambert sphere compared to the geometrical 
sphere. In our analysis we are unable to distinguish between the two cases. 
Thus, in the next section we present the results from the Lambert sphere 
case, which gives the larger companion mass.

\section{Results}

With all three cases we find that the orbit is circular and that the
eccentricity value is $e=0.0\pm$0.05. All three scenarios require
strong ellipsoidal variations to explain the phase light curve. The hybrid
model $C_{hybr}$ is significantly preferred with a confidence of 99 per cent
confidence (2.55$\sigma$) compared to $C_{therm}$, and 99.6
per cent (2.85$\sigma$) compared to $C_{refl}$. Scenarios
$C_{therm}$ and $C_{refl}$ are indistinguishable at the 1$\sigma$
level. Table \ref{tab2} shows the results of all three cases
(excluding primary transit and secondary eclipse) and
Fig. \ref{models} (top) shows the Kepler light curve and the components for the
$C_{hybr}$ scenario. In all three cases, we agree on the mass of the
unseen companion to within a few per cent, suggesting that the
technique is largely independent of the precise nature (be it thermal
or reflected) of the emission arising from the companion. Although
note that the mass constraint improves as the model becomes more
complex. Both $C_{therm}$ and $C_{refl}$ are significantly less good
than the hybrid model, suggesting that components of both thermal and
reflected light are contributing to the phase light curve. Note that we do not
give errors on $\epsilon$, $a_{\rm bol}$ and $a_{\rm g}$ in $C_{hybr}$
because they are essentially unconstrained at the 1-$\sigma$ level,
with large degeneracies between $a_{\rm g}$, $a_{\rm bol}$, and
$\epsilon$.

\begin{table}
\centering
\label{tab2}

\begin{tabular}{lrrrr}
 & $C_{refl}$ & $C_{therm}$ &  $C_{hybr}$ \\
&[Reflected] & [Thermal] & [Mixed] \\
\hline 
$M_{\rm C}$ $[M_{\rm J}]$& 8.0$\pm$1.1  & 8.3$\pm$1.0  & 8.3$\pm$0.9   \\
$\epsilon$       &                & 0.17$\pm0.17$ & 0.10  \\
$a_{\rm bol}$        &                & 0.01$+0.1$ & 0.80  \\
$a_{\rm g}$          & 0.28$\pm0.1$  &               & 0.28$\pm$0.28 \\
e [deg]          & 0.0$\pm$0.05  & 0.0$\pm$0.05         & 0.0$\pm$0.05  \\
$\chi ^{2}_{\nu}$      & 1.042          & 1.032          & 1.03   \\
 \hline
 \end{tabular}
 \caption{Model results for the three main scenarios discussed in
   the text. From left-to-right the models increase in complexity. We
   list the fitted parameters and their 1-$\sigma$ errors. In each case
   we are fitting 90 datapoints in the phase-folded light curve. For
   scenario $C_{hybr}$, the parameters shown without errors are
   essentially unconstrained by the data.}

\end{table}

The mass derived from the best fitting model of the candidate is
$M_{\rm C}=8.3 \pm 0.9 M_{\rm J}$, rather more massive than a typical
hot Jupiter, but significantly below the brown dwarf or M dwarf mass
proposed by Szabo et al. (2011). This error on $M_{\rm C}$ is the
error from fitting the model, and does not allow for the uncertainty
in the mass of the primary, rather this should be seen as a 20 per
cent error on the mass ratio, $M_{\rm C}/M_*$=0.0039. We take the mass
from Szabo et al. (2011) to be 2.05M$_\odot$ and assume an error of 10
per cent. We also consider a 0.1 per cent error in the radius ratio
$R_{\rm C}/R_*$ (Barnes et al. 2011). Accounting for both of these
leads to a final constraint on the mass of KOI-13.01, $M_{\rm C}=8.3
\pm 1.25 M_{\rm J}$.

\begin{figure}
 \includegraphics[width=80mm]{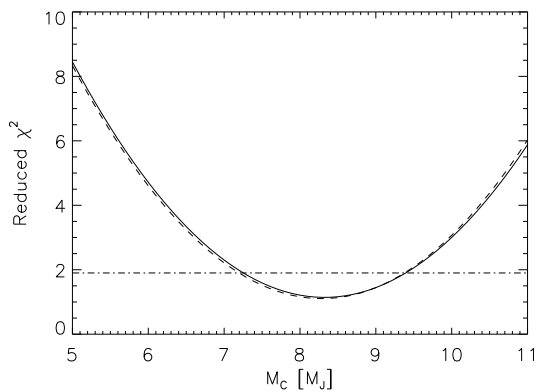} 
 \caption{$\chi_{\nu}^{2}$ vs $M_{\rm C}$ for model fits with and
   without the secondary eclipse ($C_{hybr}$, $C_{hybr}^{b}$) (solid and 
   dashed curves respectively.  The horizontal line illustrates
   the 1-$\sigma$ error level for each model.}
 \label{chi}
\end{figure}

\subsection{How hot is KOI-13.01?}

Any attempt to constrain the brightness temperature of the companion
is severely hampered by the problem of distinguishing between the
reflected and thermal components of our model, exacerbated by the
single passband. In principle, if there were no reflected light, nor
ellipsoidal variations, we could use the difference between the depth
of the secondary eclipse (star only) and the point just before ingress
to (or after egress from) the primary transit (i.e. star plus planet
night-side) for an orbit with zero eccentricity (e.g. TrES-2,
\cite{b15}). In practice, this is rather more complicated, because
very small contributions from the ellipsoidal variations, and the
marginally visible reflected light components are contributing to the
total light of the system at this point in phase. And they need to be
included in the modelling to constrain the night-side contribution to
the system flux. Mazeh et al. (2011) have ignored these effects, and
estimate the temperature of the companion to be 2600K. Our rather more
complete treatment of the light curve includes both thermal and reflected
components, and includes the secondary eclipse in a self-consistent
manner.

This then is our motivation for scenario $C_{hybr}^{b}$ introduced in
the previous Section. We added back into the light curve an additional 36 (in
eclipse) data points, while adding no additional parameters to the
model, i.e. the shape of the secondary eclipse is completely described
by $C_{hybr}$. A little surprisingly, the additional data do nothing
to improve our constraints on the nature of the secondary as
parameterized by $a_{\rm g}$, $a_{\rm bol}$ and $\epsilon$. Perhaps
this should not be so surprising, the depth of the secondary eclipse
in a single filter is unlikely to tell us very much about the ratio of
thermal to reflected emission from the planets atmosphere. Infrared
measurements could perhaps make a significant impact here, as they
could be combined with the optical passband to shed light on the
colour of the planet. We also note that our solution to the mass of
KOI-13.01 is not affected by the additional data points (see
Figure~\ref{chi}).

\section{Discussion}

The very small amplitude of the ellipsoidal variations $(\sim 7.2
\cdot 10^{-5})$ for this period, can only be explained by a body of
planetary mass. Increasing the mass of the companion would
significantly affect the amplitude of the ellipsoidal variations
(Fig. \ref{models} - bottom), therefore we are confident that the mass
of the companion KOI-13.01 is 8.3$\pm$1.25 $M_{\rm J}$ (i.e a
1$\sigma$ upper limit of 9.55$M_{\rm J}$), assuming that there are no
other contributions to the Kepler light curve that have so far been ignored in
our analysis. It is worth noting that the ellipsoidal component of the
light curve is visibly non-symmetric about phase 0.5 due to Doppler Beaming
(Figure~\ref{models}). The difference in heights is $\sim 4.3\cdot
10^{-6}$, i.e. 3 per cent of the total phase function signal (assuming
a simple model which neglects reddening).  The mass we derive for the
unseen companion to KOI-13A, places the object below the minimum mass
for deuterium burning (Spiegel et al. 2011).

One possible source of error in our estimation of the companion's
mass, is the uncertainty in the contribution of the companion A-star
to the total Kepler light curve. If we increase the contribution from the
nearby stellar component, and the flux we attribute to KOI-13A itself
decreases, the mass of the planet becomes correspondingly larger, as
the relative contribution of the ellipsoidal variations goes
up. However, the effect is rather small when compared to our other
sources of error. Specifically, a factor of $\pm 0.2$ magnitudes,
leads to a change in the mass of $\pm 0.05 M_{\rm J}$ for the
companion. If the Howell et al. (2011) delta-magnitude is used, then
the mass of KOI-13b will decrease by $\sim 20$ per cent; this
  implies the body is even more planet-like.

Another possible source of error in our treatment of the system is our
assumption that the companion behaves as a Lambert sphere. If the
correct phase function should be represented by a geometrical sphere,
then the best-fit mass of the system will actually be significantly
reduced by a systematic factor of 36 per cent, down to $M_{\rm
  C}=5.3M_{\rm J}$. See Mislis et al. (2011) for a more thorough
discussion of this issue. For this paper, we are using the Lambert 
sphere phase function, which gives the maximum $M_{\rm C}$.

The only real way to increase the mass of the unseen companion into
the brown-dwarf or stellar regime, is to add yet another unresolved
(in the Szabo et al. 2011 lucky imaging), and comparably bright,
companion to the Kepler light curve. This seems unlikely, given that the
spectral analysis of Szabo et al. (2011) is consistent with the
photometry and the model isochrones. Ultimately, the mass of KOI-13.01
can only be fully determined with high spatial resolution (to resolve
the components A and B), time-resolved spectroscopy. We estimate that
the radial-velocity amplitude for the KOI-13A system will be $K \sim$ 
870 m s$^{-1}$ for our best-fit mass. Although measuring
precise radial-velocities for a rapidly rotating and spatially-blended
A-star is not simple, the formula in Battaglia et al. (2008) suggest
that it should be eminently achievable, even with a small telescope,
given sufficient signal-to-noise. Ignoring the complications arising
from disentangling the two spectra, we find that a signal-to-noise
ratio of $~10,000$ per resolution element, should enable us to reach a
velocity precision of around 100 metre/second.

\begin{figure}
 \includegraphics[width=80mm]{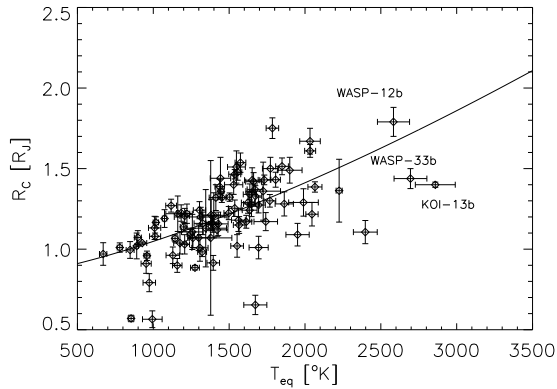} 
 \caption{$T_{eq}$ vs $R_{\rm C}$ for planets found on
   exoplanet.org. KOI-13.01 is in the middle right part of the plot. The solid line
   is the $R_{\rm C} \propto T_{eq}^{1.4}$ model from Laughlin et
   al. (2011). For the plot we have used exoplanets.org (http://exoplanets.org/).}
 \label{temper}
\end{figure}

The low amplitude of the ellipsoidal variations rules out the
suggestion from \cite{b7} that the companion could be an M star
(Fig. \ref{models}). Therefore KOI-13.01 is an exoplanet, and it is
now illustrative to consider the object in the light of other
exoplanets, especially perhaps WASP-33b which also orbits an
A-star. In the $T_{eq}$ vs $R_{\rm C}$ diagram (Fig. \ref{temper}), we
see that the current radius \citep{b17} is in good agreement with the
$T_{eq}$.  Laughlin et al. (2011) examine the radius anomaly of
exoplanets around their best fit $R \propto T^{\gamma}_{eq}$ ( with
$\gamma = 1.4$) relation, and we note that KOI-13.01, as a candidate
for one the hottest planets known, may provide a useful constraint on
our understanding of planetary structures and atmospheres.

Mazeh et al. (2011) find a mass of 4$\pm$2 and 6$\pm$3 $M_{\rm J}$
based on ellipsoidal variations and Doppler beaming effect using
$R_{\rm C}$, $R_{\star}$ from Szabo et al. 2011.  These values have
since been updated by Barnes et al. (2011). In our analysis we found
that the companion is significantly heavier, smaller and closer to the
star, than the Mazeh et al. 2011 measurement. Also, Shporer et
al. (2011) found $M_{\rm C}\sin(i)$ = 9.2$\pm$1.1 $M_{\rm J}$ based
solely on the Doppler beaming effect (also using the Szabo et al. 2011
mass). The Doppler beaming signature in the light curve is roughly an order of
magnitude smaller than the ellipsoidal variation, as Shporer et
al. (2011) show in their paper. Assuming an inclination of i=85.9, the
planetary mass from Shporer et al. is 8.1 $\leq$ $M_{\rm C}$ $\leq$
10.3. This mass value is significantly heavier than the Mazeh et
al. value (both teams are using the same algorithm and the same system
parameters), but agrees with our mass value. 

\section{Conclusions}
\label{conclusions}

We present modeling of the KOI-13.01 Kepler light curve, confirming
the planetary nature of the companion. Our analysis illustrates the
wealth of information that can be extracted from high precision light curves,
in the absence of spectroscopy.  We find that the out-of-eclipse light curve of
KOI-13.01 is dominated by two effects. The light contributed by the
planet is at the same amplitude as the ellipsoidal variations induced
in the star.

We are unable to solve for the relative contributions of thermal and
reflected emission from the planet, however we expect the planet to be
one of the hottest known, given its close proximity to an A-star, and
its large radius. The equilibrium temperature for any zero-albedo
companion to KOI-13.01 is $T_{eq} = T_{eff}\sqrt{R_{*}/2\alpha} =
2864$K, which is rather hotter than calculated for WASP-33b and
WASP-12b.  We suggest that infrared observations would help to
disentangle the thermal and reflected components of the light curve.

By modelling the light curve, we find that the mass of the planet is $M_{\rm
  C}=8.3 \pm 1.25 M_{\rm J}$, robust against any assumptions about the
albedo of the planet. We also find that the planet is orbiting in a
circular orbit ($e=0.0\pm0.05$). Our results suggest that the density of 
the planet is $\sim$3 times larger than Jupiter's density. 
This value is not surprising. There are much more dense exoplanets, 
such as XO-3b (density $\sim$6.5 times more dense than Jupiter - exoplanet.org)

We have studied the case that the system is an M-dwarf or brown-dwarf
eclipsing companion, but neither of these solutions can explain the light curve
we observe. The expected radial velocity amplitude for KOI-13 system
is $K \sim$ 870 m s$^{-1}$, which will be relatively easy to
measure despite the nature of the primary (a rapidly rotating A star)
and the contamination from the close A-star companion. It would be an
interesting exercise to search all of the Kepler light curves to look for the
signature of ellipsoidal variations as a planet detection method, even
in the absence of transits.

\section*{Acknowledgments}

This research has made use of NASA's Astrophysics Data System
Bibliographic Services. DM is supported by RoPACS, a Marie Curie
Initial Training Network funded by the European Commission’s Seventh
Framework Programme.

\bsp

\label{lastpage}

\end{document}